\documentstyle[12pt]{article}
\input math_macros

\begin{document}
 
Mermin       correctly    points  out   that   Bohr  did   not  use   the  term
``action-at-a-distance''    in  connection with  the  influence that  was under
discussion in his  exchange with EPR. Nor did the  latter authors. But Einstein
later spoke of  ``spooky   action-at-a-distance'' and  identified it as the key
point when  he said$^8$  ``One can  escape from  this [EPR]  conclusion only by
either assuming  that the measurement  of $S_1$  ((telepathically)) changes the
real situation of  $S_2$ or by denying  independent real  situations as such to
things that are spatially  separated from each  other." Bohr himself translated
the mathematical EPR argument into a physical procedure that involved ``a rigid
diaphragm with two  parallel slits,  which are very narrow  compared with their
separation, and through  each of which one particle  with given momentum passes
independently of the  other.'' An experimenter  subsequently can measure either
the  momentum or the  position  of one of  the two  particle, and  the point of
Bohr's argument was to  show that this measurement  of one particle {\it does},
within his knowledge-based,  prediction-oriented framework, influence the other
particle, thereby  upsetting the condition  ``without in any way disturbing the
system'',  which EPR  require in  order to  apply their  criterion  of physical
reality. 

Mermin's claim to have found an error in my argument hinges on his claim that I
set down conditions under which  counterfactuals can be used. But I set down no
such conditions.   I argued against  the wholesale  exclusion of counterfactual
argumentation by  giving some simple  examples. Mermin misread me if he thought
I was  requiring all   counterfactuals to be  of that  simplest  kind. Rather I
adhered throughout to the rules [modal  logic] that philosophers have developed
for specifying the meanings of counterfactual conditions.

Thus Mermin should not claim that  there was an error in my reasoning: that I
violated precepts that I myself had set forth. Rather, if he wants to limit the
meaningfulness of counterfactual  statements then he should show why the normal
rules  of   counterfactual  reasoning  [as  developed by philosophers who deal 
with this problem]  must be augmented in a way that makes my proof invalid.

Mermin's central claim is that my statement

(S): If R2 was performed and gave +, then performing R1, instead, would 
have given --.

is neither true nor false, if L1 is  performed, but  meaningless.  Mermin's 
argument
for this is that: ``I am merely noting that the statement fails to meet the
rule for  when  a   counterfactual  makes  sense.''  Later  he  says ``The
essential  ambiguity  responsible for the gap  in Stapp's  argument lies not in
LOC2  but in  the  statement  (S) to  which  Stapp  wishes to  apply  LOC2. The
counterfactual part of (S), which apparently refers only to past events  (which
is essential for the applicability of  LOC2) actually makes implicit  reference
to   future  events,   through  the    criterion  for the    validity of a
counterfactual  statement.  This implicit  future reference  does not make LOC2
{\it fail}: it renders LOC2 {\it inapplicable.}

Mermin  refers ``the  rule'' or  ``the  criterion'' for  the  meaningfulness or
validity of a counterfactual  statement. The rule he proposes was taken from my
discussion of a simple example. I will  accept this criterion as a conservative
{\it sufficient condition} for meaningfulness. This is useful because it leads,
as Mermin points out, to the  meaningfulness of (S) under condition L2. This is
important because quantum  physicists are prone to  the easy view that only the
actually   performed  experiment  has any   significance,  and hence  {\it all}
counterfactual  argumentation is  meaningless. That view  would make the entire
Bohr-EPR  exchange  meaningless,  because  the entire  discussion  is about the
effect of changing a free  choice made in  one place upon  physical reality
elsewhere, and this notion is  formulated in terms of counterfactuals. Bohr did
not challenge  EPR on the  basis of the  meaninglessness of  counterfactuals in
science.  Rather he  affirmed  that one  could regard  the choices  made by the
experimenters as  ``free'', and could talk in the  normal way about such things
{\it at the level of experimental  procedures and observable possible results}.
He took  the  tack of  redefining  the  meaning of   ``physical  reality''. 
Thus Mermin's challenge to my argument is basically dissimiliar to Bohr's
challenge to EPR.

The importance of the conservative  Stapp-Mermin criterion is that it provide a
toe-hold. By  validating (S)  under condition  L2 it  establishes that there is
something to talk about. For it establishes that Nature has some deep structure
that goes beyond what actually  happens: it says that, under the condition that
L2 is  performed later,  there is in  Nature some  connection of  what actually
happens in the  earlier region to {\it  what would have  happened} there if the
experimenter  there had made  the opposite  choice. This  derived property {\it
would have  happened} is not based on  any assumption of  determinism: it comes
out of  the  predictions of  quantum  theory  itself, in   conjunction with the
locality condition LOC1. 

Having  established  under the  conservative  criterion that  Nature has a deep
structure  that  extends beyond  the bounds  of the  single  actually occurring
situation, and relates actual happenings to {what would have happened if...} we
have made the big leap.  In exploring further this  deep structure we must make
sure that  our words  are not  just verbal  formulas:  they are  supposed to be
describing various other aspects of this deep  structure of nature.  Without 
that connection of our words to the now-establihed deep structure the whole 
project would collapse.
 
Philosophers   have  worked  out rules  [modal  logic]  for  dealing  with this
situation:  these  rules allow  our use of  words to  conform  correctly to the
natural  meaning normally  ascribed to  those  words. I adhere  to those rules,
which are completely appropriate to  the situation under consideration here,
where our purpose is to  explore properties of  nature that involve connections
among alternative possibilities that we are free to create. 

Statement (S) has a natural  meaning that is captured in modal logic. This
condition  is  formulated within  a  theoretical  context of pairs  of possible
worlds that are almost the same, but  that differ in some specified way at some
first  time   where  there is  a   significant    difference. The   theoretical
contemplation  of such  slightly differing  pairs is even more  reasonable in a
nondeterministic  quantum  context than in a  deterministic  classical context,
since within the former context one  should be able to contemplate two possible
worlds  that are the  same at  some time  except that  in one   the decay of a
radioactive nucleus has  just been detected, and  this detection will cause the
second  experiment to be  performed, whereas  in the other  possible world this
decay has  not been  detected, and the  first  experiment will  consequently be
performed. 

The meaning of (S) closely parallels  the meaning of an ordinary statement ``If
under condition A result x appears,  then under condition B result y appears''.
The essential difference  is that the extra word  ``instead'' in (S) breaks the
usual rule that truth of the first  condition (R2) allows one to conclude that
the contradictory premise (R1) of the subordinate clause is false. This change
allows one to relate the possibilities for the outcome of the two alterative
possible experiments in the way specified by the sentence. 
 
This  ``meaning''  of  statement (S)  is  strictly in  terms of a  relationship
between the possibilities for the  outcomes of alternative possible experiments
both of which are  confined to the  region R. This asserted connection between
the possibilities for the outcomes of the two alternative possible experiments
may, by virtue of the deep structure of nature, be true or false under various 
possible conditions, but they do not automatically become  ``meaningless''  
if not directly provable in some simplest way.

Given this natural meaning, the  statement (S) could conceivably remain true if
L1 is performed instead of L2. LOC2 is  just the assertion that (S) does remain
true under this substitution. This property is  plausible because  the free 
choice  between L1 and L2 does not occur  until after  everything  referred to 
in (S)  has already either happened  or not  happened: so  it is hard  to see 
how  the truth  of (S) could depend on  the later  free  choice without  there 
being  some  backward-in-time influence.

Mermin  claims  that  the  criterion  for   meaningfulness,  regarded  now as a
necessary condition, shows that LOC2  cannot be applied because the ``meaning''
of (S) ``implicitly'' involves events in L. In my earlier reply I countered his
argument by ascribing the natural meaning to (S) [i.e., the meaning in terms of
constraints on  possibilities  of outcomes]  simply as a  matter of definition,
which I am  entitled to do,  and in fact am  obliged to do  because this is the
strict meaning by virtue of which (S)  is proved to be true under condition L2.
Then I defined  LOC2 as simply asserting  that (S) held true also under 
condition L1. Then is there no question about the ``applicability'' of this
condition of LOC2: it is nothing more than the {/it assertion} that the deep
structure of nature is such as to make that connection between possibilities
of outcomes that is directly described by (S) continue to hold if the later 
free choice is to perform L1 instead of L2.   

Once it has been  established that  the deep structure in nature exists, then 
one  should adhere consistently  to the natural meaning of (S) as a 
description of a possible consequence of that structure, rather  than 
sometimes branding (S) as meaningless when, as a  description of a possible 
consequence of this deep structure that is under investigation, it has a 
natural meaning.

\end{document}